\documentclass[twocolumn,showpacs,preprintnumbers,amsmath,amssymb]{revtex4}
\usepackage{graphicx}
\usepackage{dcolumn}
\usepackage{bm}

\begin{document}

\preprint{APS/123-QED}
\title{Dimensional Crossover in Bragg Scattering from an Optical Lattice}

\author{S. Slama}
\author{C. von Cube}
\author{A. Ludewig}
\author{M. Kohler}
\author{C. Zimmermann}
\author{Ph.W. Courteille}
\affiliation{Physikalisches Institut, Eberhard-Karls-Universit\"at T\"ubingen,
\\Auf der Morgenstelle 14, D-72076 T\"ubingen, Germany}

\date{\today}

\begin{abstract}
We study Bragg scattering at 1D optical lattices. Cold atoms are confined by the optical dipole force at the antinodes of a standing
wave generated inside a laser-driven high-finesse cavity. The atoms arrange themselves into a chain of pancake-shaped layers located at
the antinodes of the standing wave. Laser light incident on this chain is partially Bragg-reflected. We observe an angular dependence of
this Bragg reflection which is different to what is known from crystalline solids. In solids the scattering layers can be taken to be
infinitely spread (3D limit). This is not generally true for an optical lattice consistent of a 1D linear chain of point-like scattering
sites. By an explicit structure factor calculation we derive a generalized Bragg condition, which is valid in the intermediate regime.
This enables us to determine the aspect ratio of the atomic lattice from the angular dependance of the Bragg scattered light.
\end{abstract}


\maketitle

Bragg scattering is a widely used method for observing and analyzing periodic structures. Introduced by von Laue and Bragg more than
80 years ago, it has become an inestimable tool in solid state physics and crystallography \cite{Wollan32}. In quantum optics the
advent of powerful laser cooling and trapping techniques has led to the realization of \textit{optical lattices}, i.e.~periodic
arrangements of ultracold atoms confined to arrays of optical potentials formed by one or more standing light waves \cite{Jessen96}.
Optical Bragg scattering from 3D optical lattices has been first investigated by Birkl \textit{et al.}~and Weidem\"uller \textit{et
al.}~\cite{Birkl95,Weidemuller95}. Bragg scattering from a 1D optical lattice has been realized recently for the first time within our
group \cite{Slama05}. At present, optical lattices play an important role in many experiments. The observation of Bloch oscillations
\cite{BenDahan96} and the realization of Mott insulators \cite{Greiner02} and Tonks-Girardeau gases \cite{Paredes01} in degenerate
atomic quantum gases are prominent examples. One-dimensional optical lattices have interesting effects on the collective behavior of
Bose-Einstein condensates \cite{Inguscio01}. Bragg diffraction represents a powerful tool for sensitively probing the properties of such
optical lattices. A method for phase-sensitive Bragg spectroscopy based on heterodyning the Bragg-reflected light with a reference
light field has recently been presented by our group \cite{Slama05}.

In this paper we show how the Bragg condition well-known from diffraction experiments with X-rays at solids has to be modified, if the
size of the crystal is limited. This is done by an explicit calculation of the structure factor. We experimentally test our model by
probing the angular dependence of the Bragg condition on a 1D optical lattice. This enables us to determine the aspect ratio of the
atomic lattice. We find that our optical lattice occupies an intermediate position between a linear chain of point-like scatterers and
a stack of extended homogeneous reflection layers reminiscent to a dielectric mirror.

\begin{figure}[ht]
        \centerline{\scalebox{0.33}{\includegraphics{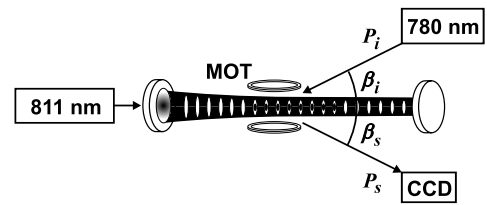}}}\caption{
            The experimental setup consists of a cavity pumped at $811~$nm and a diode laser at $780~$nm for Bragg scattering. The
            Bragg-reflected light is observed on a CCD camera.}
        \label{BraggFigSetup}
        \end{figure}

The optical layout of our experiment is shown in
Fig.~\ref{BraggFigSetup}. It consists of an optical cavity and a
setup for Bragg scattering. The cavity input coupler has a
curvature $\rho_{\mathrm{ic}}=50~$cm and a transmission
$T_{\mathrm{ic}}=0.2~\%$, and the high reflecting mirror is plane
and has a transmission $T_{\mathrm{hr}}=5\times10^{-6}$. The
measured finesse of the cavity is $4000$, and the beam diameter at
its center is $w_{\mathrm{dip}}=220~\mu$m. The light of a
titanium-sapphire laser operating at
$\lambda_{\mathrm{dip}}=811~$nm is coupled and frequency-locked to
the cavity, thus forming a standing wave with periodicity
$\frac{1}{2}\lambda_{\mathrm{dip}}=\pi/k_{\mathrm{dip}}$. The
intracavity power is $P_{\mathrm{cav}}=5~$W. Between
$N_{\mathrm{tot}}=10^5$ and $10^7$ $^{85}$Rb atoms can be loaded
from a standard magneto-optical trap into the standing wave, which
is red-detuned with respect to the rubidium $D_1$ line. From
absorption spectroscopy of the atomic cloud we roughly estimate
that about $10000$ antinodes are filled with atoms. Typically the
temperature of the cloud is on the order of a few $100~\mu$K; we
noticed in earlier experiments \cite{Kruse03,Nagorny03} that the
temperature of the cloud tends to adopt a fixed ratio with the
depth of the dipole trap. Therefore, the spatial distribution of
the atoms does not vary much with the chosen potential depth. For
the trap in this setup we measured
$k_{\mathrm{B}}T\approx0.4~U_0$. From this we derive the
\textit{rms}-size of the atomic layers,
$2\sigma_z=\frac{1}{\pi}\lambda_{\mathrm{dip}}\sqrt{k_\mathrm{B}T/2U_0}\approx115~$nm
in the harmonic approximation of the trapping potential. The
radial size is
$2\sigma_r=w_{\mathrm{dip}}\sqrt{k_{\mathrm{B}}T/U_0}\approx140~\mu$m
and the mean density can be adjusted between $n=3\times10^9$ and
$3\times10^{11}~$cm$^{-3}$.

The light used to probe the Bragg resonance is generated with a
near infrared laser diode operating at
$\lambda_{\mathrm{brg}}=780~$nm. The frequency is tuned to the Rb
$D_2$ line. The laser light is collimated to a beam waist of
$w_{\mathrm{brg}}=800~\mu$m before crossing the standing wave
under an angle of
$\beta_i=\arccos(\lambda_{\mathrm{brg}}/\lambda_{\mathrm{dip}})=15.9^\circ$.
The irradiated light intensity is, with a total laser power of
$P_i= 300~\mathrm{nW}$, far below the saturation intensity. Some
time after loading the atoms into the standing wave the Bragg
light beam is switched on. The light reflected from the atoms,
$P_s$ (several nW), is detected with a CCD camera (Sony XC55),
from which we get the intensity profile of the reflected light
beam. This allows us to determine the reflection angle $\beta_s$.

The Bragg condition follows from energy- and momentum
conservation. Our standing wave dipole trap represents a 1D
optical lattice with the lattice constant
$\mathbf{d}=\frac{1}{2}\lambda_{\mathrm{dip}}\hat{\mathbf{e}}_z=d\hat{\mathbf{e}}_z$.
The Bragg condition requires that the difference beween the
scattered and the incident wavevectors,
$\mathbf{q}\equiv\mathbf{k}_s-\mathbf{k}_i$, coincides with a
vector of the reciprocal grating $\mathbf{R}_j=j\mathbf{G}$, where
$\mathbf{G}\equiv2k_{\mathrm{dip}}\mathbf{\hat{e}}_z$. This
implies
    \begin{align}\label{EqBraggCond}
    \tfrac{1}{2}\lambda_{\mathrm{dip}}\cos\beta_i+\tfrac{1}{2}\lambda_{\mathrm{dip}}\cos\beta_s & =\lambda_{\mathrm{brg}}~,\\
    \beta_i & =-\beta_s~.\nonumber
    \end{align}
The first equation is the condition for constructive interference of light reflected from subsequent scattering planes. The second
equation arises from the fact that the radial atomic distribution is nearly homogeneous on the length scale of a wavelength. Together
the two equations~(\ref{EqBraggCond}) yield
    \begin{equation}\label{EqBraggAngle}
    \lambda_{\mathrm{dip}}\cos\beta_i=\lambda_{\mathrm{brg}}~.
    \end{equation}
The efficiency of Bragg scattering depends critically on the angle of incidence $\beta_i$. In order to probe the Bragg condition
$\beta_i$ has to be varied over the value given by~(\ref{EqBraggAngle}). Experimentally it is easier to vary the wavelength of the
lattice laser $\lambda_{\mathrm{dip}}$, while the angle of incidence is kept fixed. To resonantly enhance the Bragg scattering, which
otherwise would be neglegibly small, we tune the laser to the transition between $5S_{1/2},F=3$ and $5P_{3/2},F'=4$ at
$\lambda_{\mathrm{brg}}=780~$nm with a natural linewidth of $\Gamma_{\mathrm{brg}}/2\pi=6~$MHz. During the Bragg pulse sequence the
repumping laser of the magneto-optical trap is kept on to avoid optical pumping into the ground state $F=2$ level.

The Bragg condition~(\ref{EqBraggCond}) implies two equations.
Their claim is that for infinitely extended layers the angle of
incidence and the reflection angle are equal and their cosines sum
up to a fixed value. Both conditions are fulfilled, if the
incident beam is shone under the Bragg angle given by
(\ref{EqBraggAngle}) onto the atomic cloud. However when the angle
of incidence is misaligned from the Bragg condition, one of these
equations must be violated. Which one it is depends on the form of
the atomic cloud. For radially extended clouds, we expect the two
angles to be equal. In contrast, for long lattices (many layers)
we expect that the sum of the angles stays constant. This can be
illustrated with a calculation of the structure factor
$S(\mathbf{q})$. Its absolute square is proportional to the
scattered light intensity \cite{Coley01},
    \begin{equation}\label{GlLeistung}
    \frac{dP}{d\Omega}\propto\left|S(\mathbf{q})\right|^2=\left|\int_Vn_a(\mathbf{r})e^{i\mathbf{qr}}d^3\mathbf{r}\right|^2~,
    \end{equation}
where $n_a(\mathbf{r})$ is the atomic density within the lattice. We assume for each layer a Gaussian density distribution, which is
well fulfilled in the harmonic approximation
    \begin{equation}\label{GlDichteebene}
    n_l(\mathbf{r})=n_0\text{e}^{(-x^2-y^2)/2\sigma_r^2}\text{e}^{-z^2/2\sigma_z^2}~.
    \end{equation}
The atoms are spread over $N_s$ layers of the lattice yielding an overall density distribution, which can be expressed as a convolution
of the single site distribution with a sum of $\delta$-functions,
    \begin{equation}\label{GlDichtegitter}
    n_a(\mathbf{r})=\sum_{m=1}^{N_s}\delta(\mathbf{r}-m\mathbf{d})\star n_l(\mathbf{r})~.
    \end{equation}
Inserting (\ref{GlDichtegitter}) into (\ref{GlLeistung}) gives
    \begin{equation}\label{GlSfaktor}
    \left|S\right|^2=\left|\sum_{m=1}^{N_s}e^{im\mathbf{q}\mathbf{d}}\int_Vn_l(\mathbf{r})e^{i\mathbf{qr}}d^3\mathbf{r}
        \right|^2~.
    \end{equation}
The sum in~(\ref{GlSfaktor}) can be written as an Airy function
    \begin{equation}\label{GlAiry}
    \left|A\right|^2=\left|\sum_{m=1}^{N_s} e^{imq_zd}\right|^2=\frac{1-\cos(N_sq_zd)}{1-\cos(q_zd)}~.
    \end{equation}
Experimentally relevant is first order scattering $q_zd=2\pi$, for
which the Airy function reaches a maximum with an approximated
full width at half height of $2\delta
k_z=2\sqrt{3(5-\sqrt{5})}/N_sd\simeq2\times2.88/N_sd$. For this
approximation the cosines of the Airy function are expanded up to
sixth order and $N_s \gg 1.$ The integrals in
equation~(\ref{GlSfaktor}), one for every direction in space, are
evaluated by Gaussian functions, for example
    \begin{equation}\label{EqGaussApprox}
    \left|B(q_x)\right|^2\equiv\left|\int e^{iq_xx}e^{-x^2/2\sigma_r^2}dx\right|^2=2\pi\sigma_r^2
        e^{-q_x^2\sigma_r^2}~.
    \end{equation}
On Bragg resonance $q_x=q_y=0$ and $q_z=2k_{\mathrm{dip}}$. The
full width at half height of $|B(q_x)|^2$ and $|B(q_y)|^2$ is
given by $2\delta k_{x,y}=2\sqrt{\ln 2}/\sigma_r$. The
Debye-Waller factor $|B(q_z)|^2$ can be regarded as a constant
attenuation of the structure factor within the above calculated
range $2\delta k_z$.
        \begin{figure}[ht]
        \centerline{\scalebox{0.55}{\includegraphics{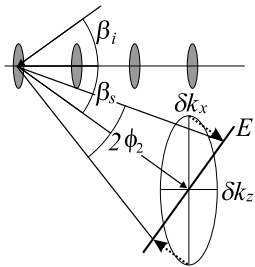}}}\caption{
            Bragg reflection in real space. The grey ellipsoids symbolize the atomic layers. Within the scattering plane the opening angle
            $2\phi_2$ of the Bragg-beam is found by projecting the widths $\delta k_x$ and $\delta k_z$ onto a plane $E$ orthogonal to the
            emission direction. The size of the angle is given by the larger of the two projections, here $\delta k_x$.}
        \label{BraggProjection}
        \end{figure}

The above calculated widths $\delta k_{x,y,z}$ correspond to a certain solid angle $\Omega$, into which the Bragg-scattered light is
emitted. The solid angle is given by $\Omega=\pi\phi_1\phi_2$, with the half opening angles $\phi_1, \phi_2 \ll 1$. The situation is
simple for the $y$-direction, which is orthogonal to the scattering plane: $\phi_1\approx\delta k_y/k_{\mathrm{brg}}$. Within the
scattering plane the widths $\delta k_x$ and $\delta k_z$ have to be projected onto a plane orthogonal to the emission direction,
as shown in figure~\ref{BraggProjection}. The opening angle $\phi_2$ is then determined by the maximum of the two projections:
$\phi_2=\max\left\{\frac{\delta k_x}{k_{\mathrm{brg}}}\cos\beta_s, \frac{\delta k_z}{k_{\mathrm{brg}}}\sin \beta_s\right\}.$ The total
result for the solid angle is therefore
    \begin{equation}\label{EqSolidAngle}
    \Omega=\pi\frac{\sqrt{\ln 2}}{\sigma_rk_{brg}^2}\max\left\{\frac{\sqrt{\ln 2}}{\sigma_r}\cos\beta_s,\frac{2.88}{N_sd}\sin\beta_s
        \right\}~.
    \end{equation}
        \begin{figure}[ht]
        \centerline{\scalebox{0.5}{\includegraphics{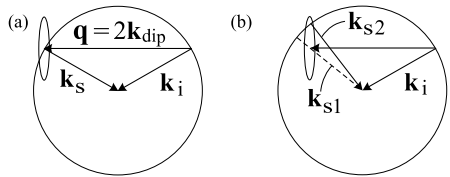}}}\caption{
            Bragg reflection in reciprocal space. The pictures show cuts through the Ewald sphere and the ellipsoid structure factor. In
            (a) the probe laser frequency and angle fulfill the Bragg condition. In (b) the lattice constant has changed. The outgoing
            wavevector $k_{\mathrm{s1}}$ has been chosen such that $\beta_s=-\beta_i$, the wavevector $\mathbf{k}_{s2}$ such that
            $\cos{\beta_s}+\cos{\beta_i}=2\lambda_{brg}/\lambda_{dip}$. The light will be emitted in the direction where the structure
            factor reaches its maximum on the intersection with the Ewald sphere.}
        \label{BraggFigScheme}
        \end{figure}

In our case $\frac{\sqrt{\ln 2}}{\sigma_r}\cos\beta_s\gg\frac{2.88}{N_sd}\sin\beta_s$, i.e.~the lattice behaves more like a linear
chain of point-like scatterers. We find $\Omega=6.6\times10^{-6}$ srad. We now calculate the direction of the emitted light beam. To
this purpose we assume that the structure factor is an ellipsoid with Gaussian profile
    \begin{align}\label{GlStrukturgauss}
    S & =S_0\exp\left(-\frac{(k \sin\beta_s-k \sin\beta_i)^2}{2\delta k_x^2}\right.\nonumber\\
    & \left.-\frac{(k\cos\beta_s-2k_{\mathrm{dip}}+k\cos\beta_i)^2}{2\delta k_z^2}\right)~,
    \end{align}
and search for the angle $\beta_s$, for which the structure factor gets largest on the intersection of the ellipsoid with the Ewald
sphere (see Fig.~\ref{BraggFigScheme}). Implicitly contained is the assumption that the scattering is elastic by setting
$k_i=k_s=k_{\mathrm{brg}}$. The part vertical to the scattering plane ($y$-direction) is omitted, because its effect on the
scattering angle is negligible. Maximization of equation~(\ref{GlStrukturgauss}), $\partial S/\partial\beta_s=0$, results in
    \begin{equation}\label{Glkbeb}
    \delta k_z^2-\delta k_x^2=\frac{\delta k_z^2\sin\beta_i}{\sin\beta_s}+\left(\cos\beta_i-\frac{2k_{\mathrm{dip}}}
        {k_{\mathrm{brg}}}\right)\frac{\delta   k_x^2}{\cos\beta_s}~.
    \end{equation}
Two limiting cases are interesting: For small aspect ratios, $\delta k_z/\delta k_x\ll 1$, we recover the Bragg condition
    \begin{equation}\label{EqSmallAspect}
    \beta_s\approx\arccos\left(2k_{\mathrm{dip}}/k_{\mathrm{brg}}-\cos\beta_i\right)~.
    \end{equation}
For large aspect ratios, $\delta k_z/\delta k_x\gg 1$, we get the second of the equations~(\ref{EqBraggCond}).
    \begin{equation}\label{EqLargeAspect}
    \beta_s\approx-\beta_i~.
    \end{equation}

The impact of a finite structure factor can be seen in experiment. There are two signatures: 1.~The reflection angle should deviate
from the values predicted by the classical Bragg condition~(\ref{EqBraggCond}). 2.~The efficiency of the Bragg scattering should
exhibit a narrow resonance upon tuning the lattice constant via $\lambda_{\mathrm{dip}}$. The width of this resonance is given by the
radial spread of the layers within the lattice. These signatures are observed in the following measurements.
        \begin{figure}[htbp]
        \centering\scalebox{0.65}{\includegraphics{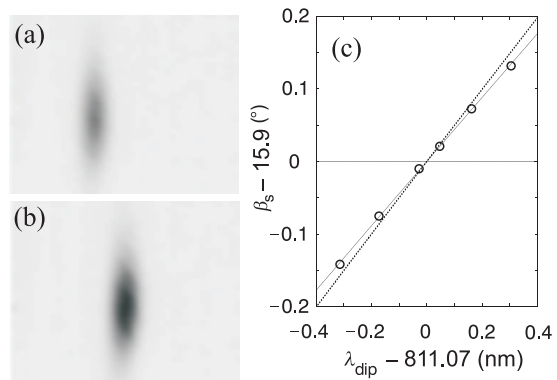}}
        \caption{
            (a),(b) CCD-pictures of the Bragg-reflected light for two different lattice constants. By fitting a Gaussian curve to the
            picture (not shown here) the center position of the beam is determined. From the displacement of the beams the relative
            emission angle of the light beams is calculated. (c) Output angle as a function of lattice constant. The measured values
            (rings) are compared to various curves: The horizontal line is expected for $\beta_s=-\beta_i$, the dotted line for
            condition~(\ref{EqSmallAspect}), and finally the solid line takes account of the finite lattice size and is a fit of
            Eq.~(\ref{EqAspectRatio}) to the data points. Fitting parameter is the aspect ratio $\zeta$. The experimental error of the
            data points lies well within the plotted circles.}
        \label{Braggpicture}
        \end{figure}

In order to detune the Bragg condition we vary the lattice
constant via the wavelength of the lattice laser. For each lattice
constant the Bragg-diffracted light is shone onto a CCD camera
(see Figs.~\ref{Braggpicture}(a),(b)). By fitting a Gaussian curve
to a horizontal cut through the image we get the center position
of the beam. The pixel size of the camera is
$l_{\mathrm{px}}=7.4~\mu\mathrm{m}$. Together with the distance
between camera and atomic cloud we determine the relative emission
angle of the Bragg-beam for various lattice constants.
Figure~\ref{Braggpicture}(c) shows that the reflection angle
varies with the lattice constant, thus $\beta_s\neq-\beta_i$.
Furthermore, the reflection angle slightly deviates from
equation~(\ref{EqBraggAngle}) (dotted line), but follows the
generalized Bragg condition~(\ref{Glkbeb}) (solid line). The
findings demonstrate that our system is far from the assumption of
infinitely extended layers. By introducing the aspect ratio
$\zeta=\frac{\delta k_z^2}{\delta k_x^2}$ equation~(\ref{Glkbeb})
is rewritten as
    \begin{equation}\label{EqAspectRatio}
    \frac{1}{\zeta-1}\left(\zeta\frac{\sin\beta_i}{\sin\beta_s}+\frac{\cos\beta_i-2\frac{k_{dip}}{k_{brg}}}{\cos\beta_s}\right)=1~.
    \end{equation}
        \begin{figure}[ht]
        \centerline{\scalebox{0.7}{\includegraphics{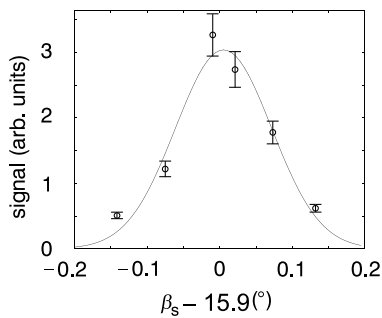}}}
        \caption{
            Intensity of the reflected light against emission angle. The data points (circles) are fitted with a Gaussian curve. The
            acceptance angle of the Bragg-reflection equals the width of this fit. The error of the data points is due to shot-to-shot
            fluctuation and lies in the 20\% range.}
        \label{ampfit}
        \end{figure}

By fitting the data from figure~\ref{Braggpicture}(c) with equation~(\ref{EqAspectRatio}) (fitting parameter $\zeta$) we get an aspect
ratio of $\zeta=0.01$. This means that the ratio of the width to the length of the lattice $\frac{2\sigma_r}{N_sd}=2.9\%$. The radial
size is approximately $2\sigma_r\approx 140~\mu$m, we therefore calculate a lattice length of $N_sd=4.8~$mm. This means that about
$N_s=12000$ layers take part in the Bragg scattering process, which exceeds our rough guess of the lattice length by 20\%. The
comparatively narrow radial atomic distribution implies a self-adjustment of the Bragg condition to the lattice constant,
i.e.~$\cos{\beta_s}+\cos{\beta_i}$ is automatically kept constant. Still, small deviations from this situation (dotted line in
Figure~\ref{Braggpicture}(c)) are experimentally seen.

The tolerance for the acceptance angle is equal to the divergence of the outgoing beam $2\phi_2$. As can be seen from the calculation
of the output solid angle (Eq.~(\ref{EqSolidAngle})) $2\phi_2=2\sqrt{\ln2}/\sigma_rk_{br}=0.17^\circ$. To compare this value with the
experimental data, the intensity of the reflected light extracted from figure~\ref{Braggpicture}(a),(b) is plotted against the
emission angle. These data are then fitted by a Gaussian curve (see Fig.~\ref{ampfit}). The full width at half maximum of this fit is
$2\delta\beta=0.16^\circ$, which agrees very well with the theoretical value of $2\phi_2$.

In conclusion, we studied Bragg-reflection at a 1D optical lattice. From a structure factor calculation we deduce a generalized Bragg
condition which is valid not only in the solid state physics limit of infinitely spread scattering layers, but also for the case where
boundary effects play a role, with the extreme limit of a linear chain of localized scatterers. By comparing the theoretical results
with measurements of the emission angle of the Bragg-reflected light we show that our optical lattice is very close to the latter case.
This case is characterized by the fact that it is hard to detune the light scattering away from the Bragg condition, because the
emission angle self-adjusts to the Bragg condition. This can have undesired consequences for the probing of one-dimensional photonic
bandgaps predicted to appear in optical lattices as a consequence of multiple reflections between subsequent layers
\cite{Deutsch95,Coevorden96}. Indeed their signatures are most conveniently probed at a detuned Bragg angle, such that the band edge
spectrally lies outside the resonance frequency of the atomic transition. In this way, the region close to the atomic transition, where
multiple reflection competes with diffuse light scattering \cite{Birkl95} can be avoided. This is not possible, if the lattice is close
to being a chain of point scatterers. Ways to overcome this problem include the choice of larger beam diameter for the dipole trap
(eventually by confining the atoms in higher-order TEM modes \cite{Kruse03}), smaller Bragg angles and longer lattices.

\bigskip

We acknowledge financial support from the Landes\-stiftung Baden-W\"urt\-temberg.

\end{document}